# First-principles thermoelasticity of bcc iron under pressure

Xianwei Sha and R. E. Cohen

Carnegie Institution of Washington, 5251 Broad Branch Road, NW, Washington DC 20015, USA

We investigate the elastic and isotropic aggregate properties of ferromagnetic bcc iron as a function of temperature and pressure, by computing the Helmholtz free energies for the volume-conserving strained structures using the first-principles linear response linear-muffin-tin-orbital method and the generalized-gradient approximation. We include the electronic excitation contributions to the free energy from the band structures, and phonon contributions from quasi-harmonic lattice dynamics. All the elastic moduli increase with increasing pressure, and decrease with increasing temperature. The isotropic aggregate sound velocities obtained based on the calculated elastic moduli agree well with available ultrasonic and diamond-anvil-cell data. Birch's law, which assumes a linear increase in sound velocity with increasing atomic density, fails for bcc Fe under extreme conditions. First-principles linear response lattice dynamics is shown to provide a tractable approach to examine the elasticity of transition metals at high pressures and high temperatures.

## I. Introduction

Information on the influences of pressure and temperature on the elastic moduli and related aggregate properties of single crystals plays an essential role in predicting and understanding the interatomic interactions, strength, mechanical stability, phase transition mechanisms and dynamical response of materials. During the past several decades, considerable experimental efforts have been devoted to examine the elasticity of bcc iron, as well as its temperature and pressure dependences.[1-10] There are also some first-principles data available, but these calculations only focus on elasticity at zero temperature, without any thermal effects included.[11, 12] Here we have performed first-principles quasiharmonic lattice dynamics study to examine the elastic moduli of bcc Fe with pressures and temperatures, using the full-potential linear response linear-muffin-tin-orbital (LMTO) method.

## II. Theoretical methods

For a cubic crystal, the three elastic moduli $C_{11}$, $C_{12}$ and $C_{44}$ fully describe its elastic behavior. $C_{11}$ and $C_{12}$ can be determined from the bulk modulus K and shear constant $C_s$

$$K=(C_{11}+2C_{12})/3 \qquad (1)$$

$$C_s=(C_{11}-C_{12})/2 \qquad (2)$$

In order to make direct comparisons to the ultrasonic experimental measurements, one should use the adiabatic bulk modulus $K_s$[13]

$$K_s=(1+\alpha\gamma T)*K_T \qquad (3)$$

where α is the thermal expansion coefficient, γ is the Grüneisen parameter, and T is the temperature. The isothermal bulk modulus $K_T$ was determined according to the Vinet equation of state (EoS).[14, 15] We recently reported first-principles thermal EoS properties for ferromagnetic bcc Fe, including α, γ and $K_T$ as a function of temperature and pressure,[16] from which we obtained the adiabatic bulk modulus $K_s$ using Eq. (3).

For cubic crystals, the shear modulus $C_s$ describes materials resistance to shear deformation across the (110) plane in the [1$\bar{1}$0] direction, and $C_{44}$ is the resistance to shear deformation across the (100) plane in the [010] direction. We obtained both the shear moduli by shearing the cubic lattice at constant volumes.[17] The elastic moduli we present are those that appear in the equations of motion and directly give sound velocities.[18-20] The following tetragonal strains are applied to obtain $C_s$

$$\varepsilon = \begin{pmatrix} \delta & 0 & 0 \\ 0 & \delta & 0 \\ 0 & 0 & (1+\delta)^{-2} - 1 \end{pmatrix} \qquad (4)$$

where δ is the strain magnitude. The Helmholtz free energy of the strained structure F(δ) is related to δ as:

$$F(\delta) = F(0) + 6C_s V \delta^2 + O(\delta^3) \qquad (5)$$

with V as the volume, and F(0) as the free energy for the unstrained structure.

In a similar way, $C_{44}$ is calculated by applying the volume-conserving orthorhombic strain:

$$\varepsilon = \begin{pmatrix} 0 & \delta & 0 \\ \delta & 0 & 0 \\ 0 & 0 & \delta^2/(1-\delta^2) \end{pmatrix} \qquad (6)$$

Here the free energy is:

$$F(\delta) = F(0) + 2C_{44} V \delta^2 + O(\delta^4) \qquad (7)$$

For many metals and alloys, the Helmholtz free energy F can be accurately separated as[13]

$$F(V,T,\delta)=E_{static}(V,\delta)+F_{el}(V,T,\delta)+F_{ph}(V,T,\delta) \tag{8}$$

$E_{static}(V,\delta)$ is the energy at zero temperature, $F_{el}(V,T,\delta)$ is the thermal free energy arising from electronic excitations, and $F_{ph}(V,T,\delta)$ is the phonon contribution. We obtain both $E_{static}(V,\delta)$ and $F_{el}(V,T,\delta)$ from first-principles calculations directly, and use linear response lattice dynamics to examine the lattice vibrational contribution in the quasi-harmonic approximation. Our computational approach is based on the density functional theory and density functional perturbation theory, using multi-κ basis sets in the full potential LMTO method.[21, 22] The induced charge densities, the screened potentials and the envelope functions are represented by spherical harmonics up to $l_{max}=6$ within the non-overlapping muffin-tin spheres surrounding each individual atom, and by plane waves in the remaining interstitial region with cutoff corresponding to the 16×16×16 fast-Fourier-transform grid in the unit cell of direct space. The **k**-space integration needed for constructing the induced charge density is performed over the 16×16×16 grid. We use the Perdew-Burke-Ernzerhof (PBE) generalized-gradient approximation (GGA) for the exchange and correlation energy.[23] When calculating $F_{el}(V,T)$, we assume temperature-independent eigenvalues for given lattice and nuclear positions.[16, 24, 25] We determine the dynamical matrix for a set of irreducible **q** points at the 8×8×8 reciprocal lattice grid. The perturbative approach is employed for calculating the self-consistent change in the potential.[26, 27] Careful convergence tests have been made against **k** and **q** point grids and many other parameters. See Ref. 16 for more computational details.

We calculate the elastic moduli at six different volumes 65, 70, 75, 79.6, 85 and 90 bohr$^3$/atom, and apply several different volume-conserving strain magnitudes varying from 1 to 3% at each given volume. For each configuration we perform first-principles linear response LMTO calculations to obtain the band structure and lattice dynamics information, and then computed the Helmholtz free energies for temperatures from 0 K to 2000 K in intervals of 250 K. We derived the elastic moduli $C_s$ and $C_{44}$ by fitting the calculated Helmholtz free energies at each given volume and temperature to Eqs. (5) and (7), respectively. We obtain the related pressure information at the given volume and temperature from the first-principles thermal EoS.[16]

## III. Results and Discussions

We present the calculated static elastic moduli of ferromagnetic bcc Fe as a function of volume (pressure) in Fig. 1 and Table 1. Several sets of experimental moduli have been published, measured by inelastic neutron scattering,[9] synchrotron x-ray diffraction,[8] low-frequency torsional,[10] and ultrasonic techniques such as rectangular parallelepiped resonance (RPR)[7] and phase comparison method.[6] At both the ambient and high pressures, the first principles calculations generally overestimate the elastic moduli. One of the major reasons for the differences here is due to the errors in the thermal equation of state. GGA approximation usually leads to smaller equilibrium volumes for bcc Fe.[11, 12, 16, 28] The agreement between the calculated moduli and the experiment is better if compared at constant volumes, as shown in Fig. 1. A couple of first-principles GGA calculations were previsously performed using the plane-wave VASP code to examine the elasticity of bcc Fe at zero temperature,[11, 12] which underestimated $C_{12}$ and $C_{44}$, but overestimated $C_{11}$. There are several aspects to account for the differences between the current and pervious theoretical calculations: the current calculation use the full potential LMTO method, and earlier calculations used ultrasoft pseudopotentials[12] and the projector-augmented-wave (PAW) method[11], respectively; both the previous calculations used the PW91 GGA functional,[29] while current calculations used the more recent PBE functional;[23] different Eos formulations might have been used to obtain the bulk modulus; and the equilibrium volumes at zero temperature are slightly different.

Although the pressure dependences of the elastic moduli of bcc Fe were studied using ultrasonic methods almost forty years ago,[1, 5] earlier studies were generally limited to very low pressures (~1GPa). Only within the past decade, new experimental techniques such as synchrotron x-ray diffraction[8] and inelastic neutron scattering[9] have been applied to examine the elastic properties at much higher pressures. All the calculated elastic moduli show a strong increase with the increase of pressure, except for $C_s$ at very high pressures. The softening of $C_s$ indicates the dynamic instability of bcc structure, consistent with earlier theoretical predictions.[28, 30] Despite the offset in the values, the calculated slopes of the elastic moduli with respect to the volume shows good agreements with experiment. The agreement between current and earlier PAW calculations[11] are good, except at large volumes. Some of the discrepancies might be attributed to the different GGA functional used. The large differences for $K_s$ at large volumes mainly come from the different equation of state formulations. Caspersen et al. used the Murnaghan EoS, and we employ the Vinet EoS. Cohen et al. examined the accuracy of various EoS formulations, and found the Vinet equation to be the most accurate.[31] We also compared with the full potential linear-augmented-plane-wave (LAPW) calculated bulk modulus using the third order Birch-Murnaghan EoS.[28] Although the LAPW and PAW calculations used the same PW91 GGA functional, the LAPW bulk modulus agrees better with current studies at large volumes.

For volumes less than ambient equilibrium value, the anisotropy ratio $A=C_{44}/C_s$ shows a strong increase with increasing pressure. This is different from bcc Ta, where A first decreases for P<100GPa, and then increases with further increase in pressure.[17]

Ultrasonic techniques have been widely used to examine the temperature dependence of the elastic moduli for single-crystal bcc Fe at ambient pressure.[4, 6, 7, 32] In Fig. 2 we show the calculated moduli $C_{11}$, $C_{12}$ and $C_{44}$ as functions of temperature at pressures from 0 to 40 GPa in intervals of 10 GPa, in comparison to these experimental data. Although the current first principle calculations overestimated $C_{11}$, both $C_{12}$, $C_{44}$ and the calculated temperature dependences of the elastic moduli show good agreements with the experiments at ambient pressure. The calculated moduli decrease with temperature in a quite linear manner at all pressures, consistent with earlier ultrasonic measurements by Leese and Lord Jr.[4] At temperatures above 750 K, the ultrasonic measured moduli show a more rapid drop than the initial linear slope, which might be associated with the effects of changing spin order on the elastic properties at high temperatures.[7] Dever reported that the magnetic contributions to $C_{11}$, $C_s$ and $C_{44}$ at Debye temperature $T_\theta$ (473 K) are 6%, 22%, and 4%, respectively[6]. Isaak and Masuda found that magnetic contributions account for 28%, 44% and 12% to the change in the $C_{11}$, $C_s$ and $C_{44}$ moduli from $T_\theta$ to the Curie temperature Tc (1043 K).[7] Although we performed spin-polarized total-energy calculations to study the properties for ferromagnetic bcc Fe, we have not included the contributions from magnetic fluctuations to the free energies and elastic moduli. This might be the main reason for the bigger differences between the calculations and the experiment as approaching the Curie temperature.

The temperature dependence of elastic moduli is often expressed as $(1/C_{ij})(dC_{ij}/dT)$. At ambient conditions, our calculations give the following values for bcc iron:

$$\frac{1}{C_{11}}\frac{dC_{11}}{dT} = -2.1 \times 10^{-4} K^{-1}$$

$$\frac{1}{C_{44}}\frac{dC_{44}}{dT} = -1.6 \times 10^{-4} K^{-1} \qquad (9)$$

$$\frac{1}{C'}\frac{dC'}{dT} = -2.2 \times 10^{-4} K^{-1}$$

Where $C'=(C_{11}+C_{12}+2C_{44})/2$. These are in good agreement with ultrasonic data by Leese and Lord Jr,[4] -2.52, -1.90 and $-2.20\times10^{-4} K^{-1}$ for $C_{11}$, $C_{44}$ and C', respectively.

Sound velocities in solids are related to the elastic moduli according to the Christoffel equation.[17] For cubic polycrystalline sample, the average isotropic shear modulus G can be determined from single crystal elastic moduli according to the Voigt-Reuss-Hill scheme[13]

$$G^V = \frac{1}{5}(2C_s + 3C_{44})$$

$$G^R = [\frac{1}{5}(\frac{2}{C_s} + \frac{3}{C_{44}})]^{-1} \tag{10}$$

$$G^H = \frac{1}{2}(G^V + G^R)$$

We show the temperature dependences of the calculated adiabatic bulk modulus $K_s$ and Hill-averaged G at different pressures in Fig. 3. Compared to the experiments, first principle calculations overestimated both $K_s$ and G, which can be largely attributed to the errors in the EoS. The calculated temperature dependences show good agreements between the theory and the ultrasonic experiment by Leese and Lord Jr.[4] The ultrasonic data of Dever[6] and Isaak and Masuda[7] show a more rapid nonlinear drop at high temperatures, which might be associated with the degree of ferromagnetic ordering. Low-frequency torsional measurements gave much smaller values for G at high temperatures, 29 and 38 GPa at frequencies of 0.01 and 1 Hz at $T_C$, respectively.[10] These values are around 14-29 GPa less than the ultrasonic measurements, and the differences might be attributed to the significant viscoelastic relaxation in bcc Fe at high temperature.[7] The torsional oscillation and creep tests at seismic frequencies and high temperatures reveal intense viscoelastic relaxation in bcc Fe.[10]

The three isotropically averaged aggregate sound velocities can be derived from bulk modulus K and shear modulus G

$$v_P = [(K + \frac{4}{3}G)/\rho]^{1/2}$$

$$v_B = (K/\rho)^{1/2} \tag{11}$$

$$v_S = (G/\rho)^{1/2}$$

Where $v_P$, $v_B$ and $v_S$ are the compressional, bulk and shear sound velocities, respectively, and $\rho$ is atomic density.

For [110] wave propagation direction in a cubic lattice, the longitudinal mode is

$$\rho v^2 = (C_{11} + C_{12} + 2C_{44})/2 \qquad (12)$$

and the two transverse modes are

$$\rho v^2 = (C_{11} - C_{12})/2 = C_s$$
$$\rho v^2 = C_{44} \qquad (13)$$

We show the calculated sound velocities of bcc Fe as a function of atomic density in Fig. 4, which agree within 3% with the ultrasonic[1, 5] and recent inelastic X-ray scattering[33] experimental data. The velocities at a given volume agree much better with the experiment then the moduli at a given pressure due to two major reasons: the errors in the thermal equation of state; and the fact that the sound velocities are the square root of the bulk and shear moduli. We show the calculated shear wave velocities as a function of atomic density at three different temperatures in Fig. 5. When the atomic density varies from 7.5 to 9 g/cm$^3$, the computed values closely obey Birch's Law, which predicts that the velocity of each material is linear with atomic density. The linear slope of the Birch's plots shows an increase with increasing temperature, similar to what has been observed in hcp Fe in recent inelastic x-ray scattering experiment.[34] However, with further compression, the computed velocities show a strong deviation from Birch's Law, and the effects are more pronounced at high temperatures. The large deviation occurs at high pressures (P> 25GPa), where bcc is still dynamically stable, but not thermodynamically stable.[16] At room temperature, bcc Fe transforms to hexagonal-close-packed (hcp) structure at ~ 11GPa.[35] Recent experiments show that the compressional and shear wave velocities of hcp Fe at high pressures and temperatures could not be fitted to Birch's law either.[34]

## IV. Conclusions

The elasticity and sound velocity of bcc Fe are presented from first-principles linear response calculations. Generally the calculated moduli are in fairly good agreements with ultrasonic, inelastic neutron scattering and x-ray diffraction measurements. However, there are some systematic shifts in moduli and the thermal equation of state from experiment, so that an improved density functional for ferromagnetic bcc Fe is desirable. Elastic moduli normally show an increase with increasing pressure and a quite linear decrease with temperature. The temperature and pressure dependences of the calculated moduli agree with experiment. The isotropic aggregate sound velocities are obtained based on the calculated elastic moduli, which show good agreements with available ultrasonic and diamond-anvil-cell data. The isotropic wave velocities follow the Birch's law only when the pressure is less than 25 GPa, with an increase in the slope of the Birch's lines with increasing temperatures.

**ACKNOWLEDGEMENTS**

We thank S. Y. Savrasov for kind agreement to use his LMTO codes and many helpful discussions. This work was supported by DOE ASCI/ASAP subcontract B341492 to Caltech DOE w-7405-ENG-48. Computations were performed on the Opteron Cluster at the Geophysical Laboratory and ALC cluster at Lawrence Livermore National Lab, supported by DOE and the Carnegie Institution of Washington.

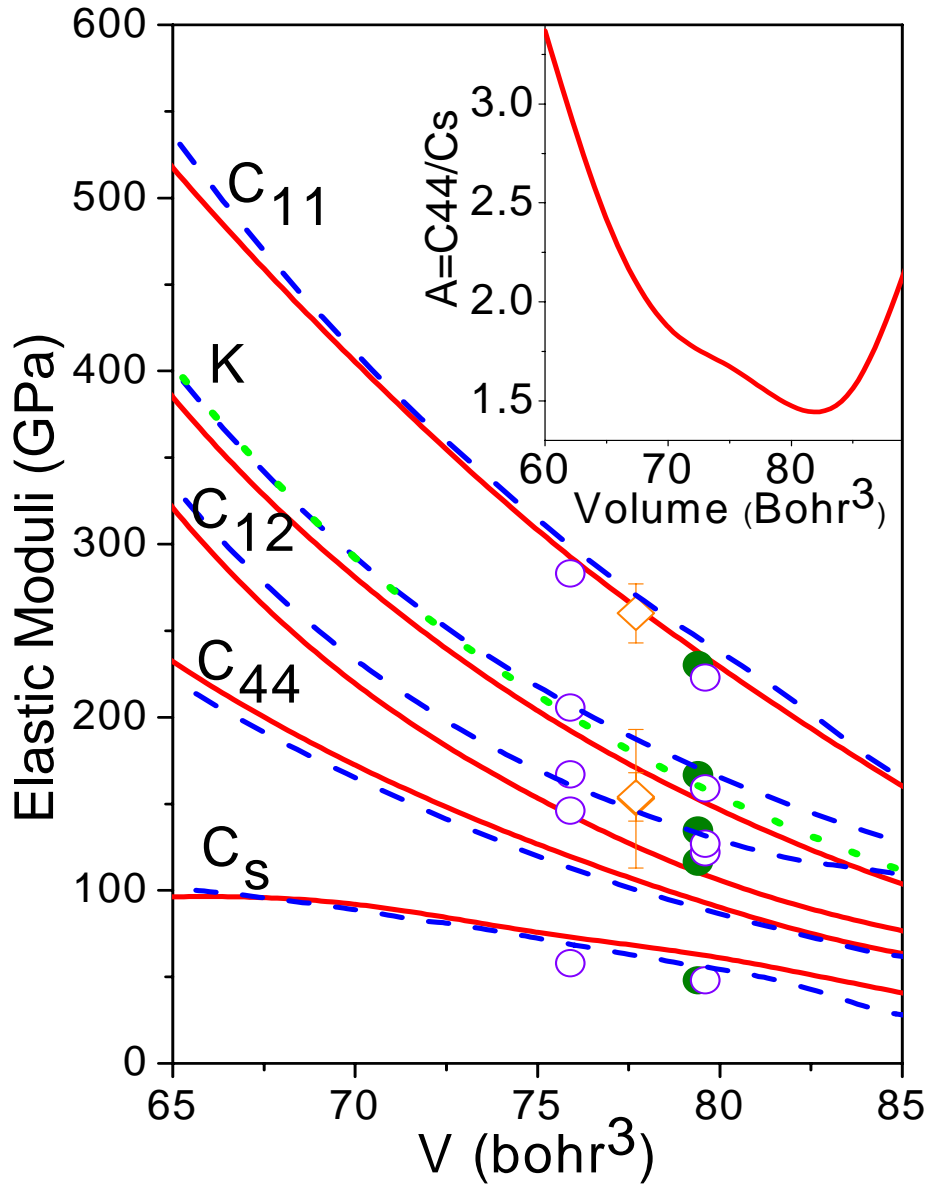

Fig. 1. The calculated static elastic constants for ferromagnetic bcc Fe as a function of atomic volume (solid lines) at ambient temperature. Open circles, filled circles and open diamonds with error bars are experimental data from inelastic neutron scattering (Ref. 9), ultrasonic (Ref. 5) and X-ray diffraction (Ref. 8) measurements, respectively. Zero-temperature theoretical PAW moduli (Ref. 11) and LAPW bulk modulus (Ref. 27) are shown in dashed and dotted lines. The inset shows the volume dependence of the anisotropy ratio.

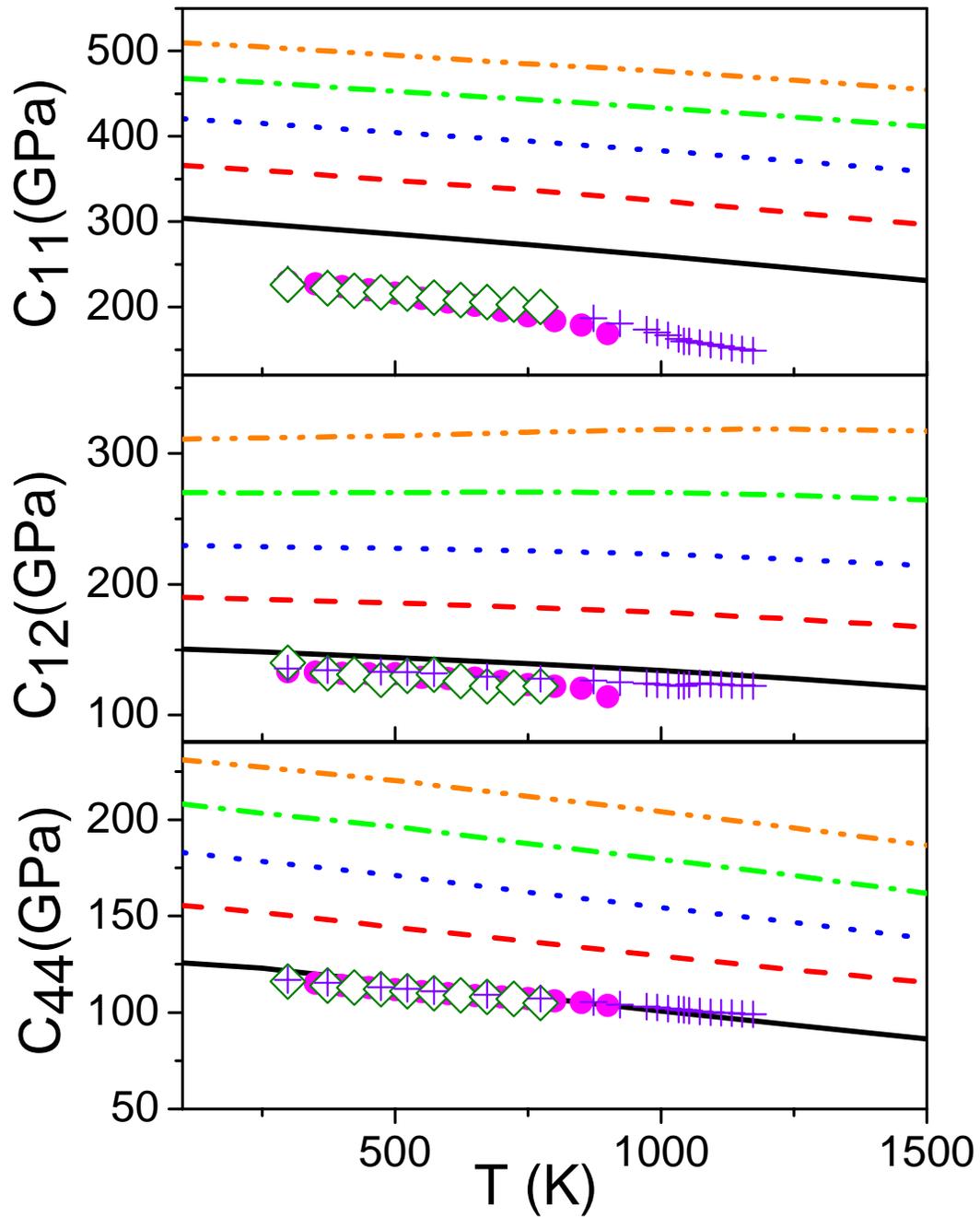

Fig. 2. The calculated temperature dependences of the elastic moduli for ferromagnetic bcc Fe, at pressures from 0 (lowest curve) to 40 GPa (uppermost curve) with 10 GPa internal. Experimental data at ambient pressure from ultrasonic measurements (cross, Ref. 6; open diamonds, Ref. 4; filled circles, Ref. 7) are also presented.

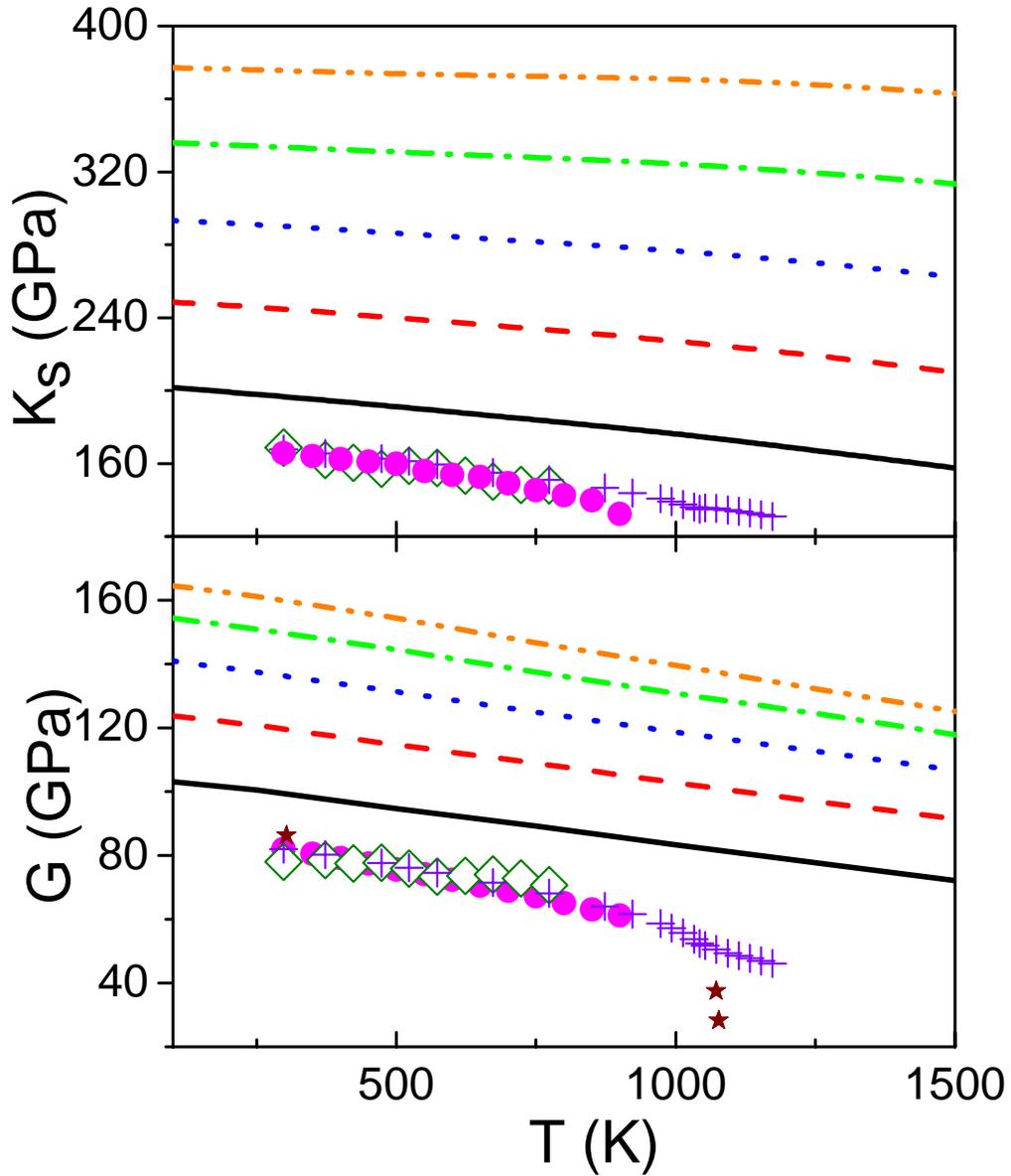

Fig. 3. The adiabatic bulk modulus $K_s$ and isotropic shear modulus G of bcc Fe as a function a temperature, at pressures from 0 (lowest curve) to 40 GPa (uppermost curve) with 10 GPa internal. Experimental data at ambient pressure from ultrasonic (cross, Ref. 6; open diamonds, Ref. 4; filled circles, Ref. 7) and low-frequency torsional measurements ( filled stars, Ref. 10) are also plotted.

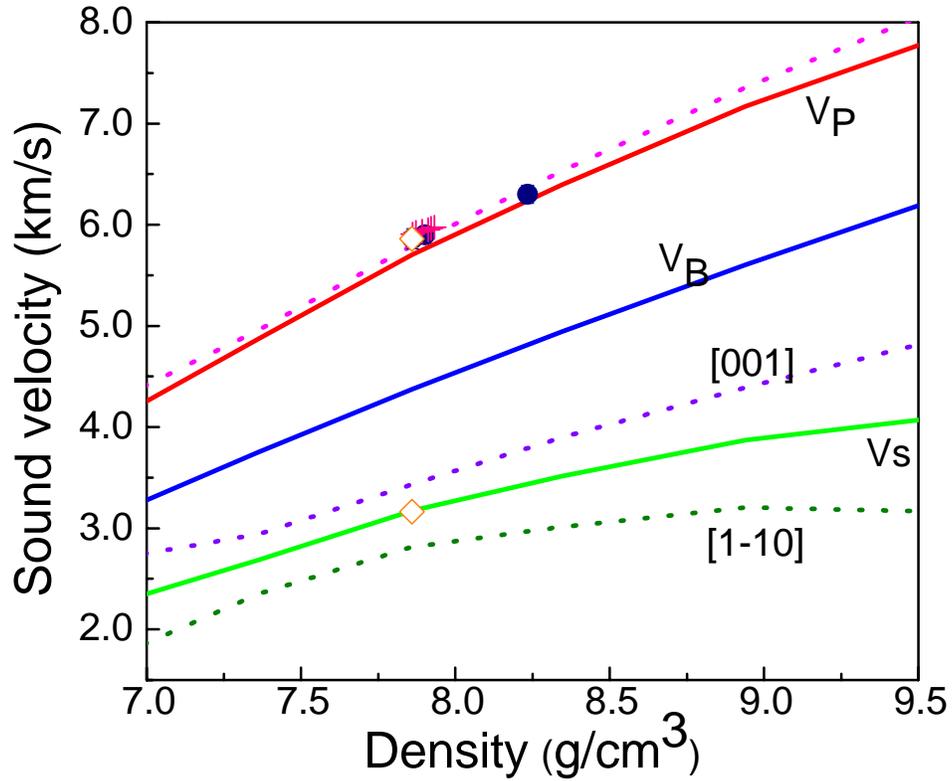

Fig. 4. Sound velocities of bcc Fe at ambient temperature calculated from elastic constants. The isotropic aggregate velocities are shown by solids lines, with $v_P$, $v_B$ and $v_S$ standing for the compressional, bulk and shear sound velocities, respectively. Dotted lines represent the longitudinal and two transverse sound velocities in the [110] direction, with brackets showing the polarization of the shear waves. Experimental isotropic compressional and shear velocities, measured by ultrasonic (open diamonds, Ref. 1; filled circles, Ref. 5) and inelastic X-ray scattering (Cross, Ref.33) are also shown.

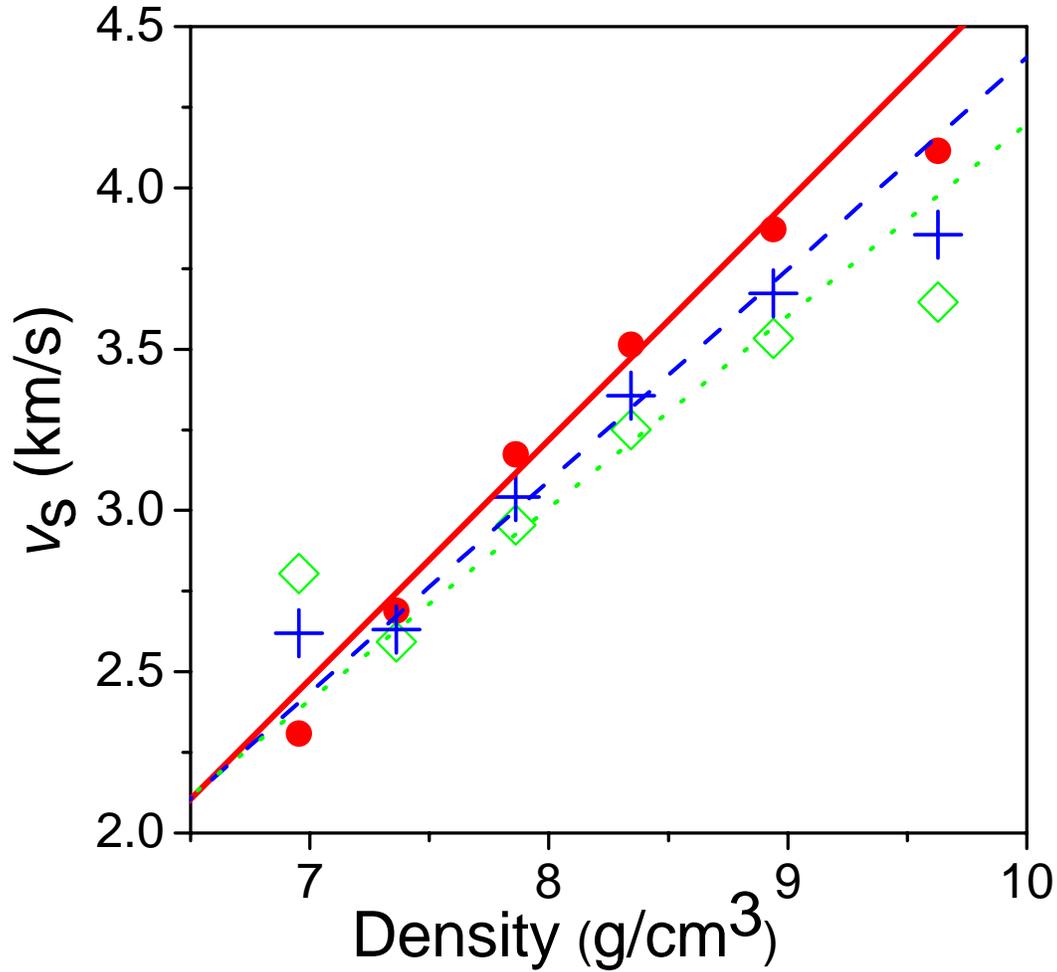

Fig. 5. Compressional wave velocity for bcc Fe as a function of atomic density. Birch law (linear dependence of the longitudinal acoustic sound velocity with the atomic density, shown as solid, dashed and dotted lines for 250, 1000 and 2000 K results, respectively) clearly fails when the atomic density is less than 7.5 g/cm$^3$, showing by the significant discrepancies from the computed data (filled circles, 250K; cross, 1000K; and open diamonds, 2000K). The slope of the Birch lines shows an increase with temperature.

TABLE I. The static elastic, bulk and isotropic shear moduli of ferromagnetic bcc Fe. Moduli are given in units of GPa, temperature T are given in Kelvin, and volume V are given in bohr$^3$/atom.

|  | T | V | $C_{11}$ | $C_{12}$ | $C_{44}$ | $K_s$ | G |
|---|---|---|---|---|---|---|---|
| **0 GPa** | | | | | | | |
| LMTO-GGA | 0 | 75.1 | 303 | 150 | 126 | 201 | 103 |
| PAW-GGA [11] | 0 | 77.1 | 271 | 145 | 101 | 187 | 84 |
| PP-GGA [12] | 0 | 76.9 | 289 | 118 | 115 | 175 | 102 |
| ultrasonic [1] | 4 | 79.0 | 243.1 | 138.1 | 121.9 | 173 | 87 |
| LMTO-GGA | 250 | 75.4 | 297 | 148 | 123 | 198 | 100 |
| ultrasonic [4] | 300 | - | 226 | 140 | 116 | 169 | 78 |
| ultrasonic [6] | 300 | 79.5 | 232.2 | 135.6 | 117 | 168 | 82 |
| RPR [7] | 300 | 79.7 | 230.5 | 133.3 | 116.3 | 166 | 82 |
| neutron [9] | 300 | - | 223 | 127 | 122 | 159 | 84 |
| **4.6 GPa** | | | | | | | |
| LMTO-GGA | 250 | 73.8 | 326 | 167 | 137 | 220 | 110 |
| X-Ray [8] | 300 | - | 260 | 154 | 153 | 189 | 100 |
| **9.8 GPa** | | | | | | | |
| LMTO-GGA | 250 | 72.1 | 360 | 188 | 152 | 245 | 120 |
| neutrons [9] | 300 | - | 283 | 167 | 146 | 206 | 101 |